\documentclass[journal]{IEEEtran}

\usepackage{amsmath,amssymb,amsfonts}
\usepackage{bm}
\usepackage[compress]{cite}
\usepackage{graphicx}
\usepackage{amsmath}
\usepackage{algorithm}
\usepackage{algpseudocode} 
\usepackage{array} 
\usepackage{enumitem}
\usepackage{booktabs} 

\usepackage{subcaption} 
\usepackage{siunitx}  

\usepackage[caption=false,font=footnotesize,labelfont=sf,textfont=sf]{subfig}
\usepackage{textcomp}
\usepackage{stfloats}
\usepackage{url}
\usepackage{verbatim}

\usepackage[dvipsnames]{xcolor}
\usepackage{diagbox}
\usepackage{multirow}
\ifCLASSINFOpdf
\else
\fi

\begin{document}

\title{\huge{Payload-Based Sensing for Beam Management:\\ Experimental Validation on a mmWave Testbed}}

\author{Yunxin~Li, Fan~Liu,~\IEEEmembership{Senior Member,~IEEE,} Hany~Assasa, Haoqiu~Xiong, \\
Raquel Marina Noguera Oishi,
and~Sofie~Pollin,~\IEEEmembership{Senior Member,~IEEE}
\thanks{Yunxin Li, Hany Assasa, Haoqiu Xiong, Raquel Marina Noguera Oishi, and Sofie Pollin are with the Department of Electrical Engineering (ESAT), KU Leuven, 3001 Leuven, Belgium.}
\thanks{Fan Liu is with the National Mobile Communications Research Laboratory, Southeast University, Nanjing 210096, China.}
\thanks{Corresponding author: Fan Liu (e-mail: fan.liu@seu.edu.cn).}
}

\maketitle
\begin{abstract}
Integrated sensing and communication (ISAC) has emerged as a promising paradigm for future 6G systems, enabling joint environment perception and data transmission. In mmWave systems, sensing-assisted beam management can reduce beam training overhead and improve mobility support. This paper presents an experimental validation of sensing-assisted beam management using standard-compliant 5G NR signals. The proposed approach exploits the communication payload signal for sensing, eliminating the need for dedicated sensing waveforms or additional pilot resources. A mmWave testbed is developed to demonstrate the feasibility of the proposed method under practical hardware constraints and realistic propagation conditions. By extracting range information from echo signals and leveraging geometric inference and prediction, proactive beam switching is achieved without periodic beam sweeping. Experimental results show that centimeter-level ranging accuracy can be obtained, and the proposed scheme maintains comparable communication performance while reducing signaling overhead, thereby improving the spectral efficiency. These results demonstrate the feasibility of payload-based sensing-assisted beam management on practical mmWave hardware and provide experimental insights for its further development.
\end{abstract}


\IEEEpeerreviewmaketitle

\vspace{-0.5cm}
\section{Introduction}

\IEEEPARstart{R}{ecent} advances in wireless systems have led to the emergence of integrated sensing and communication (ISAC) as a key capability of future networks, where radio systems are expected to provide not only data connectivity but also native environment perception. This vision has already been incorporated into the evolution of wireless standards, where the IMT-2030 framework identifies ISAC as one of the six usage scenarios for 6G~\cite{ITURM2160}, and ongoing activities in 3GPP and ETSI are exploring sensing-related functionalities and system architectures~\cite{3gpp_tr22837,3gpp_tr38765,ETSI_GR_ISC_001}. By sharing spectrum, hardware, and waveform resources, ISAC can improve spectral and energy efficiency while reducing the signaling overhead associated with separate communication and sensing operations. Beyond these resource-efficiency benefits, it also provides networks with environmental awareness that can support functions such as localization, mobility management, and beam control.

Motivated by these advantages, a large body of prior work has studied sensing-assisted communication from analytical and simulation perspectives~\cite{liu2020radar,xu2024isac,du2022integrated}, exploiting sensing information to support beam management~\cite{li2024frame}. In parallel, another line of research has explored the use of implicit channel information embedded in received communication signals to optimize beamforming or precoding without explicit channel estimation~\cite{attiah2022deep,huang2024capacity}. In particular, mmWave systems are attractive for ISAC due to their large bandwidth and directional beamforming capability, which enable fine range resolution and highly directional transmission. However, practical mmWave implementations face significant hardware challenges: analog phased arrays and limited RF chains make direct angular sensing difficult, while large bandwidths require high speed DAC/ADC operation and substantial baseband processing. In addition, synchronization and phase impairments become more pronounced at high carrier frequencies. These constraints motivate the question of whether high resolution range estimates can be exploited to support beam tracking when direct angular estimation is limited. Since such effects are often abstracted away in theoretical studies, experimental validation is essential to assess the feasibility of sensing-assisted communication under practical hardware constraints.

Experimental ISAC prototypes have gradually evolved from shared-hardware feasibility toward tighter waveform and system integration. Early common-aperture and mmWave measurement platforms demonstrated that radar and communication functions can operate simultaneously on practical hardware~\cite{mccormick2017simultaneous,kumari2020mimo}. 
Building on this foundation, later prototypes explored waveform integration by embedding information into sensing transmissions. Representative examples include spatial modulation-based designs~\cite{ma2021spatial} and OFDM joint waveform experiments examining the tradeoff between communication reliability and sensing performance~\cite{xu2022experimental}. More recent works have reused communication platforms for sensing, including massive MIMO platforms that extract radar information from channel measurements~\cite{sakhnini2022near}, networking ISAC testbeds that rely on reference signals for sensing~\cite{ji2023networking}, and real time OFDM prototypes evaluated in controlled static scenarios~\cite{zhou2023prototype}.

Despite the significant progress in experimental ISAC systems, most existing prototypes still rely on dedicated sensing signals, or specially designed pilots to achieve sensing function. As a result, additional time-frequency resources are consumed, limiting the level of integration and reducing communication efficiency. Moreover, conventional beam tracking based on received signal measurements is inherently reactive, since beam updates are triggered after link quality degradation or after explicit beam sweeping. To address this gap, this paper presents an experimental validation of sensing-assisted beam management using standard-compliant 5G NR signals on a real time mmWave testbed. The main contributions of this paper are summarized as follows:

\begin{itemize}
    \item A real-time 60 GHz mmWave testbed is developed to validate sensing-assisted communication using standard-compliant NR payload signals, without dedicated sensing resources or additional pilot allocation.

    \item A range-based proactive beam switching scheme is implemented by integrating payload-based range estimation, geometric inference, and Kalman filtering, enabling beam prediction under the hardware constraints of analog beamforming.

    \item Experimental results demonstrate centimeter-level cross-range ranging accuracy and comparable communication performance to the conventional SSB-sweeping baseline. By avoiding periodic beam sweeping, the proposed scheme further improves effective throughput under the considered overhead assumptions.
\end{itemize}

\vspace{-0.3cm}
\section{System Model}
\label{sec:systemmodel}

We consider a mmWave ISAC scenario where a gNB and a UE utilize 5G NR OFDM waveform for simultaneous communication and sensing. The gNB transmits a resource grid $\mathbf{S} = [\mathbf{s}_0, \mathbf{s}_1, \dots, \mathbf{s}_{L-1}] \in \mathbb{C}^{M \times L}$ consisting of $M$ subcarriers and $L$ symbols. A co-located receiver at the gNB captures back-scattered echoes, while the UE performs downlink decoding and acts as a primary reflector. The received signals for communication and sensing can be expressed as:
\begin{equation}
\begin{cases} 
\mathbf{Y}_\text{c} = \mathbf{H}_\text{c} \mathbf{S} + \mathbf{Z}_\text{c} \\
\mathbf{Y}_\text{s} = \mathbf{H}_\text{s} \mathbf{S} + \mathbf{Z}_\text{s} 
\end{cases}
\end{equation}
where $\mathbf{Z}_\text{c}$ and $\mathbf{Z}_\text{s}$ represent additive complex Gaussian noises. The communication channel $\mathbf{H}_c$ is modeled as a combination of a dominant Line-of-Sight (LoS) path and multiple scattering components. The sensing channel $\mathbf{H}_\text{s}$ characterizes the monostatic propagation environment, capturing the direct Tx-Rx leakage and the reflections from $K$ paths (the UE and environmental scatterers): 
\vspace{-1em}
\begin{equation}
\mathbf{H}_\text{s}
=
\underbrace{\alpha_0 \mathbf{\Lambda}(\tau_0)}_{\text{leakage}}
+
\underbrace{\sum_{k=1}^{K}
\alpha_k
\mathbf{\Lambda}(\tau_k)}_{\text{target and scatterers}}
\end{equation}
where $\alpha_k$ and $\tau_k$ denote the complex reflection coefficient and propagation delay of the $k$th path, respectively. The diagonal matrix $\mathbf{\Lambda}(\tau) = \mathrm{diag} ([1, e^{-j2\pi \Delta f \tau}, \dots, e^{-j2\pi (M-1) \Delta f \tau}]^T)$ represents the linear phase shift across subcarriers caused by the delay.

The gNB estimates the sensing channel response as $\tilde{\mathbf{H}}_\text{s} = \frac{1}{L} \mathbf{Y}_\text{s} \mathbf{S}^H$. To extract range information, an $N_{\text{IFFT}}$-point IFFT is applied to the diagonal elements of $\tilde{\mathbf{H}}_\text{s}$ with zero-padding. Letting $\tilde{p}$ denote the peak index corresponding to the UE reflection in the resulting delay profile, the measured range is:
\begin{equation}
\tilde{r} = \frac{\tilde{p} c}{2 N_{\text{IFFT}} \Delta f}
\end{equation}
where $c$ is the speed of light and $\Delta f$ is the subcarrier spacing.

\vspace{-0.3cm}
\section{Range-Based Beam Tracking and Proactive Beam Switching}
\label{sec:tracking}

\begin{figure}[!ht]
  \centering
  \includegraphics[width=0.85\linewidth]{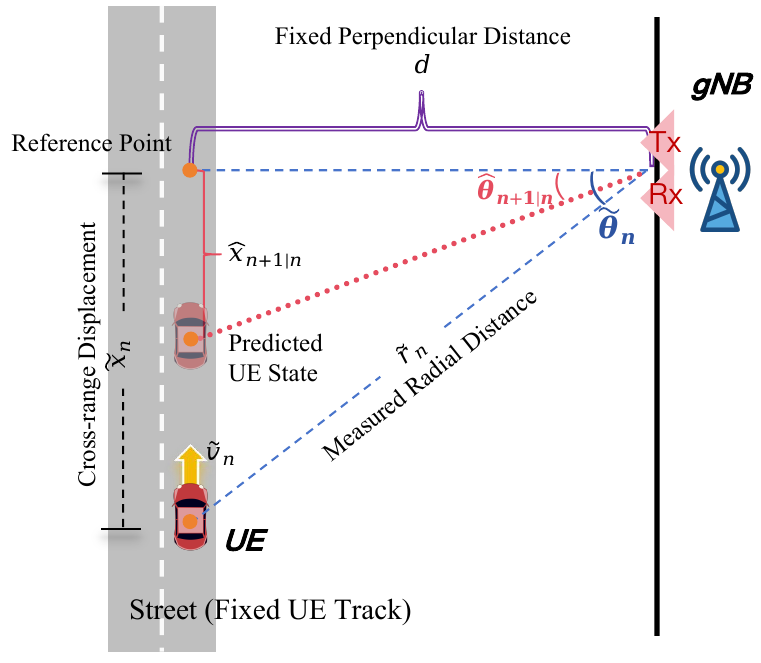}
  \caption{Roadside deployment geometry for range-based beam tracking.}
  \label{fig:scenario}
\end{figure}

We consider a street deployment where the gNB is located beside a straight UE trajectory and its antenna boresight is orthogonal to the track, as shown in Fig.~\ref{fig:scenario}. The perpendicular distance between the track and the gNB is fixed and denoted by $d$. While this geometry does not cover arbitrary UE trajectories, it represents a controlled roadside access scenario. Although the current sensing configuration does not provide explicit angle or velocity measurements, the geometric state of the UE can still be inferred from the sequence of range observations and exploited for beam tracking and beam switching.

Let $\tilde{r}_n$ denote the measured radial distance at the $n$th observation instant. The corresponding cross-range from the reference point is obtained as
$\tilde{x}_n = \sqrt{\tilde{r}_n^2-d^2}$, and the geometric angle with respect to the gNB is
$\tilde{\theta}_n = \arctan\!\left(\frac{\tilde{x}_n}{d}\right)$. Based on the sequence of range
observations, the UE motion state is tracked over time and further used to predict its angular position. This predicted angle is then used to proactively determine whether beam
switching should be triggered before communication quality degrades.

\vspace{-0.3cm}
\subsection{Range-Based Beam Tracking}

To improve robustness against measurement noise and enable motion prediction, a Kalman filter is adopted to track the one-dimensional UE motion based along the predefined track. The state vector at the $n$th observation instant is defined as $\mathbf{\xi}=[x_n,v_n]^T$, where $x_n$ and $v_n$ denote the cross-range displacement and velocity of the UE, respectively.

Since the observation interval $\Delta T$ between two adjacent measurements is short, the velocity of the UE is assumed to remain approximately constant within this interval. Accordingly, the state evolution model and the corresponding measurement model can be compactly expressed as
\begin{equation}
\left\{\begin{array}{l}
\text {State Evolution Model: }
\mathbf{\xi}_{n+1} = \mathbf{G}\mathbf{\xi}_n + \mathbf{w}_n, \\
\text {Measurement Model: } 
\tilde{x}_n = \mathbf{C}\mathbf{\xi}_n + z_n,
\end{array}\right.
\end{equation}
where
\begin{equation}
\mathbf{G}
=
\begin{bmatrix}
1 & \Delta T \\
0 & 1
\end{bmatrix},
\qquad
\mathbf{C}
=
\begin{bmatrix}
1 & 0
\end{bmatrix},
\end{equation}
$\mathbf{w}_n$ and $z_n$ represent zero-mean Gaussian noises caused by model approximation and sensing uncertainty.

Based on the standard Kalman filtering procedure, the filtered state estimate at the $n$th observation instant is denoted by $\hat{\mathbf{\xi}}_n$, and the predicted state for the next observation instant is written as
\begin{equation}
\hat{\mathbf{\xi}}_{n+1|n}
=
\mathbf{G}\hat{\mathbf{\xi}}_n.
\end{equation}
Accordingly, the predicted cross-range is given by the first element of $\hat{\mathbf{\xi}}_{n+1|n}$, denoted by $\hat{x}_{n+1|n}$. The corresponding predicted angle can then be expressed as
\begin{equation}
\hat{\theta}_{n+1|n}
=
\arctan\!\left(\frac{\hat{x}_{n+1|n}}{d}\right).
\end{equation}

\vspace{-0.5cm}

\subsection{Proactive Beam Switching}

Let $\mathcal{B}=\{\theta_i^{\mathrm c}\}_{i=1}^{N_{\mathrm B}}$ denote the beam codebook, where $\theta_i^{\mathrm c}$ is the center angle of the $i$th candidate beam and $N_{\mathrm B}$ is the total number of beams in the codebook. At each observation instant, the predicted angle $\hat{\theta}_{n+1|n}$ is compared with the angular region of the currently serving beam.

If $\hat{\theta}_{n+1|n}$ remains within the coverage region of the current beam, the beam is maintained. Otherwise, the gNB proactively switches to the beam whose coverage region contains the predicted angle. This beam selection rule can be written as

\begin{equation}
i^\star
=
\arg\min_i
\left|
\theta_i^{\mathrm{c}} - \hat{\theta}_{n+1|n}
\right|.
\end{equation}

In this way, beam switching is triggered according to the predicted geometric state of the UE rather than waiting for communication quality degradation. Although the final beam is still selected from the discrete analog codebook, range sensing provides a finer motion estimate for deciding when to switch. For an angular spacing $\Delta\theta$, the cross-range granularity is approximately $d\Delta\theta$ near boresight, whereas the large mmWave bandwidth enables centimeter-level range estimation. Thus, the proposed method improves the switching decision rather than the beam codebook resolution.

The proposed scheme is particularly applicable to locally straight road segments, where the perpendicular distance remains approximately constant over short tracking intervals. Small geometric uncertainties may be tolerated due to the finite angular coverage of practical beams. However, significant lateral deviations or turning maneuvers may affect angular prediction and beam switching accuracy. Extending the scheme to arbitrary trajectories would require more general tracking models and additional spatial information.

\vspace{-0.2cm}

\section{Testbed Implementation and Experimental Evaluation}

\vspace{-0.2cm}

\subsection{Experimental Setup}

\begin{figure}[!ht]
  \centering
  \includegraphics[width=\linewidth]{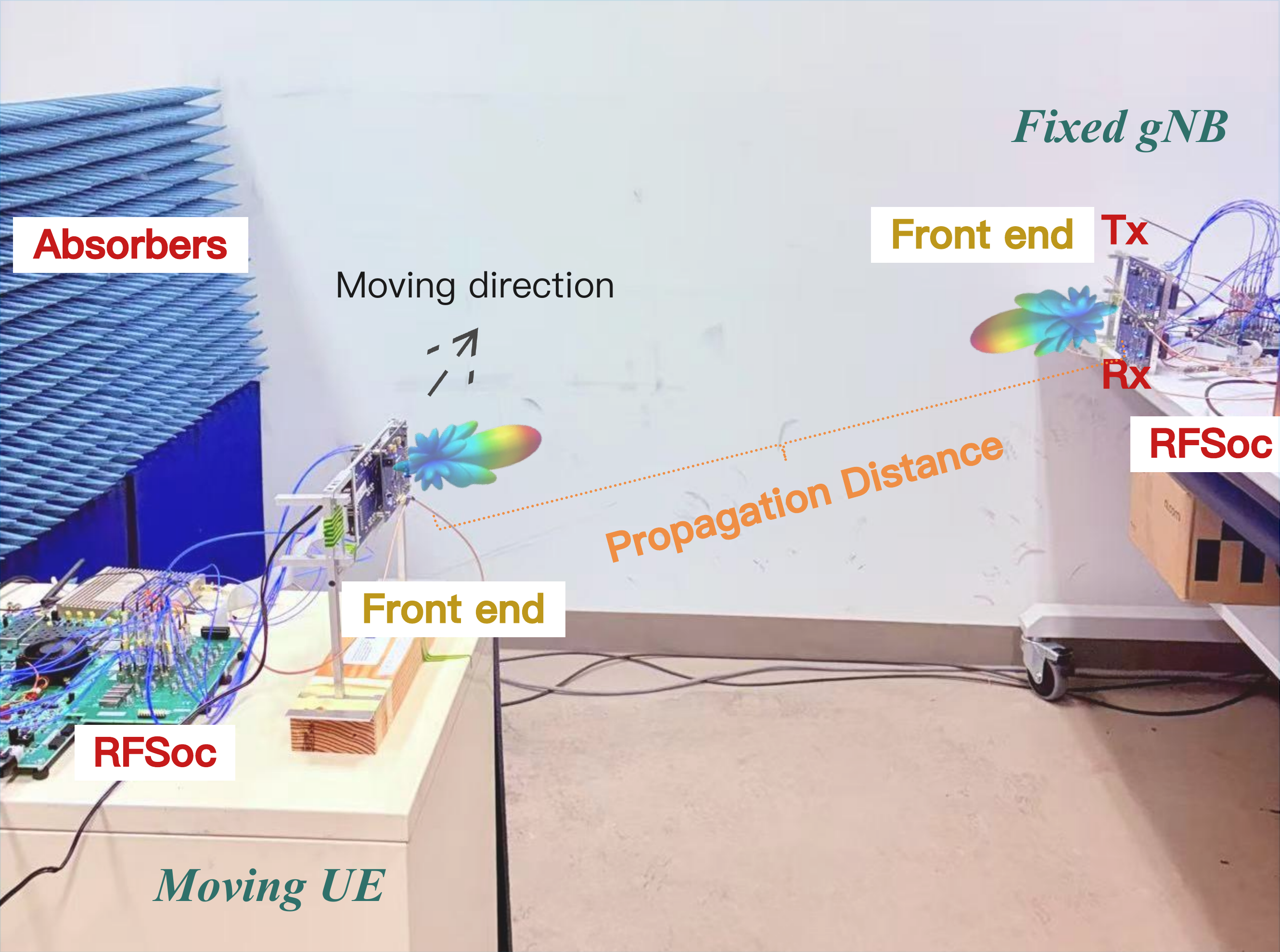}
  \caption{Laboratory setup of the mmWave ISAC testbed.}
  \label{fig:testbed}
\end{figure}

The proposed ISAC framework is validated using a wideband mmWave testbed operating at $60~\mathrm{GHz}$. The platform integrates a Xilinx Zynq UltraScale+ RFSoC for baseband processing and data acquisition with Pharrowtech $8\times8$ phased array modules for analog beamforming. As illustrated in Fig.~\ref{fig:testbed}, the gNB utilizes two co-located RF front-ends driven by a shared reference clock to ensure phase coherence for monostatic sensing. The UE, positioned at a perpendicular distance $d = 1~\mathrm{m}$, acts as both the communication user and the primary radar target.

Due to the single-RF chain analog architecture, the received signals from individual antenna elements are combined before digital processing, preventing direct high resolution AoA estimation. Consequently, the proposed scheme relies on range measurements and known geometry for beam prediction, while beam selection is limited to the predefined analog codebook. In addition, Doppler extraction requires coherent accumulation over multiple sensing snapshots and is further constrained by capture memory. The experiment therefore focuses on whether single-snapshot range estimation, enabled by the large mmWave bandwidth, is sufficient for the range-based beam tracking scheme in Section~\ref{sec:tracking}. The measurement sequence consists of $N_{\text{obs}}=25$ discrete UE positions with $5~\text{cm}$ spacing; consecutive observations are treated as snapshots separated by $\Delta T=20~\text{ms}$, corresponding to an equivalent along-track speed of $2.5~\text{m/s}$. The key experimental parameters are summarized in Table~\ref{tab:testbed_params}.

\begin{table}[!ht]
    \caption{Key experimental parameters}
    \label{tab:testbed_params}
    \normalsize
    \centering
    \begin{tabular}{p{1.7cm} p{1.7cm} p{1.7cm} p{2cm}} 
    \hline
    \hline
    Parameter & Value & Parameter & Value \\
    \hline
    $f_c$ & $60~\mathrm{GHz}$ 
    & $f_s$ & $3.52~\mathrm{Gsps}$ \\
    $BW$ & $1.73~\mathrm{GHz}$ 
    & $\Delta f$ & $3.44~\mathrm{MHz}$ \\
    $N_{\text{FFT}}$ & $1024$ 
    & $N_{\text{RB}}$ & $42$ \\
    $N_{\text{IFFT}}$ & $4096$ 
    & $N^{\text{B}}$ & $5$ \\
    $Q_{\mathrm M}$ & $2$ (QPSK) 
    & $N_{\text{payload}}$ & $12312~\mathrm{bits/slot}$ \\
    $N_{\text{T}}^{\text{gNB}},N_{\text{R}}^{\text{gNB}}$ & $64\;(8\times8)$
    & $N_{\text{T}}^{\text{UE}},N_{\text{R}}^{\text{UE}}$ & $64\;(8\times8)$ \\
    $d$ & $1~\mathrm{m}$ 
    & $x_n$ & $0.3$--$1.5~\mathrm{m}$ \\
    \hline
    \hline
    \end{tabular}
\end{table}

\vspace{-0.4cm}
\subsection{NR Signal Configuration and Beam Codebook}

The transmitted waveform follows the NR OFDM payload structure. In the experiments, the gNB transmits a Physical Downlink Shared Channel (PDSCH) waveform occupying $42$ resource blocks, corresponding to $504$ active subcarriers out of a $1024$-point FFT. The occupied bandwidth is approximately $1.73~\mathrm{GHz}$, leading to a nominal range resolution of $8.7~\mathrm{cm}$. QPSK modulation is adopted to ensure robust transmission under practical hardware impairments.

In the experiments, the transmitted waveform follows an NR OFDM payload structure with embedded Demodulation Reference Signal (DMRS) and Phase Tracking Reference Signal (PTRS) for reliable communication. Specifically, DMRS are inserted at two predefined OFDM symbol positions within each slot and occupy every other subcarrier across the allocated bandwidth for channel estimation and coherent demodulation. PTRS are transmitted with a higher time density, appearing in every OFDM symbol, while being sparsely distributed in frequency by occupying every other resource block to compensate for common phase error and residual phase noise in the mmWave front end. Importantly, no dedicated sensing waveform is introduced. The proposed scheme reuses the same NR payload waveform for sensing, without allocating dedicated sensing signals or additional time-frequency resources.

Analog beamforming is performed using a predefined beam codebook. In the considered setup, five azimuth beams are used, while the elevation angle is fixed at $0^\circ$. The corresponding beam indices and steering angles are summarized in Table~\ref{tab:beam_codebook}.

\begin{table}[!ht]
\centering
\caption{Azimuth beam codebook used for analog beamforming}
\label{tab:beam_codebook}
\small
\renewcommand{\arraystretch}{1.2}
\begin{tabular}{c|ccccc}
\hline\hline
Beam index $i$ & 1 & 2 & 3 & 4 & 5 \\
\hline
$\theta_i^{\mathrm c}$ [$^\circ$] & -48 & -36 & -24 & -12 & 0 \\
\hline\hline
\end{tabular}
\end{table}

\vspace{-0.4cm}

\subsection{SSB-Based Communication-Only Baseline}

To benchmark the proposed sensing-assisted beam switching scheme, a conventional communication-only baseline based on NR synchronization signal block (SSB) sweeping is implemented on the same testbed. In this baseline, the gNB sequentially transmits SSBs using different beam indices from the predefined beam codebook. For each candidate beam, the received beam quality is evaluated using the synchronization-signal reference signal received power (SS-RSRP). Specifically, the SS-RSRP is computed as
\begin{equation}
\mathrm{SS\text{-}RSRP}
=
10\log_{10}\!\left(
\frac{1}{N_{\mathrm{SS}}}
\sum_{n=1}^{N_{\mathrm{SS}}}
\left|X_{\mathrm{SS}}[n]\right|^2
\right),
\end{equation}
where $X_{\mathrm{SS}}[n]$ denotes the received Secondary Synchronization Signal (SSS) resource element and $N_{\mathrm{SS}}$ is the number of such elements used for averaging.

In the experiments, the SS-RSRP values corresponding to all candidate beams are measured and compared, and the beam with the highest SS-RSRP is taken as the preferred serving beam. Therefore, beam switching in the baseline is driven purely by communication measurements and occurs only after explicit beam sweeping and power comparison, making the procedure inherently reactive.

\vspace{-0.4cm}
\subsection{Experimental Results}

We next evaluate the performance of the proposed sensing-assisted scheme in comparison with the communication-only baseline. The results focus on beam switching behavior and the resulting communication performance.

\begin{figure}[!ht]
  \centering
  \includegraphics[width=\linewidth]{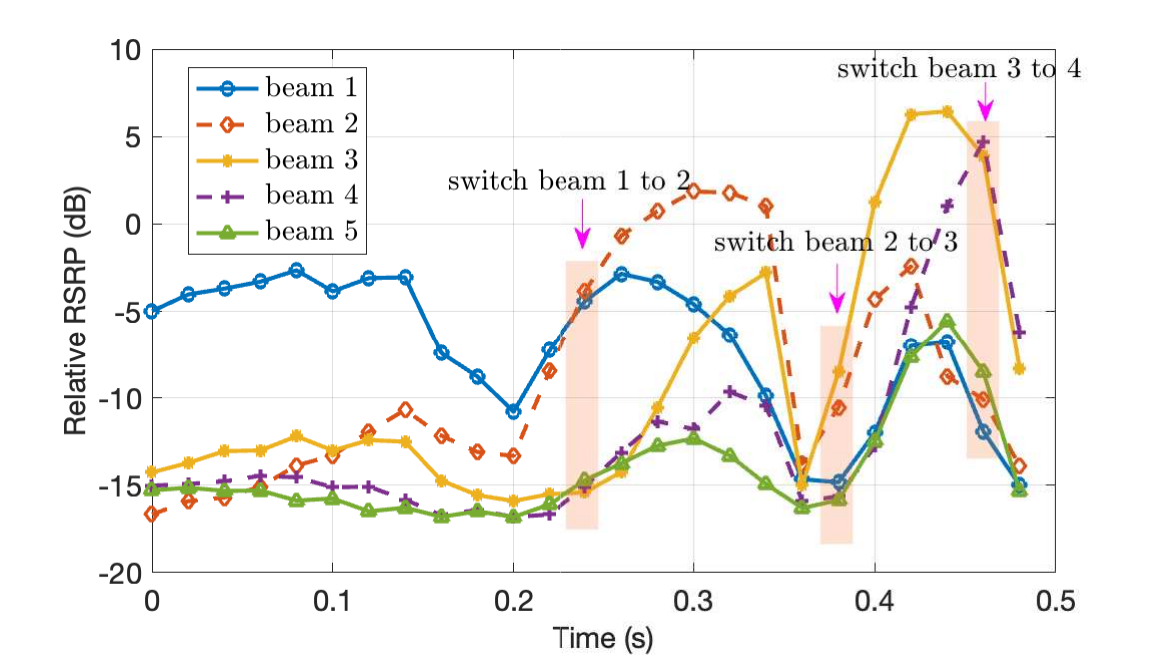}
  \caption{Relative received signal strength of the communication-only baseline.}
  \label{fig:rsrp_baseline}
\end{figure}

Fig.~\ref{fig:rsrp_baseline} illustrates the relative SS-RSRP of the communication-only baseline as the UE moves along the predefined track. Due to the directional nature of analog beamforming, each beam provides a limited spatial coverage. As the UE traverses the track, it enters and then exits the coverage of each beam, resulting in a characteristic increase followed by a decrease in the corresponding SS-RSRP. As the UE moves across adjacent beam directions, the received power of a neighboring beam eventually exceeds that of the currently serving beam, which triggers beam switching. The shaded regions indicate the observation instants at which such switching events occur. These results show that beam tracking in the baseline is achieved through periodic beam sweeping and reactive comparison of received signal strength.

\begin{figure}[!ht]
  \centering
  \includegraphics[width=\linewidth]{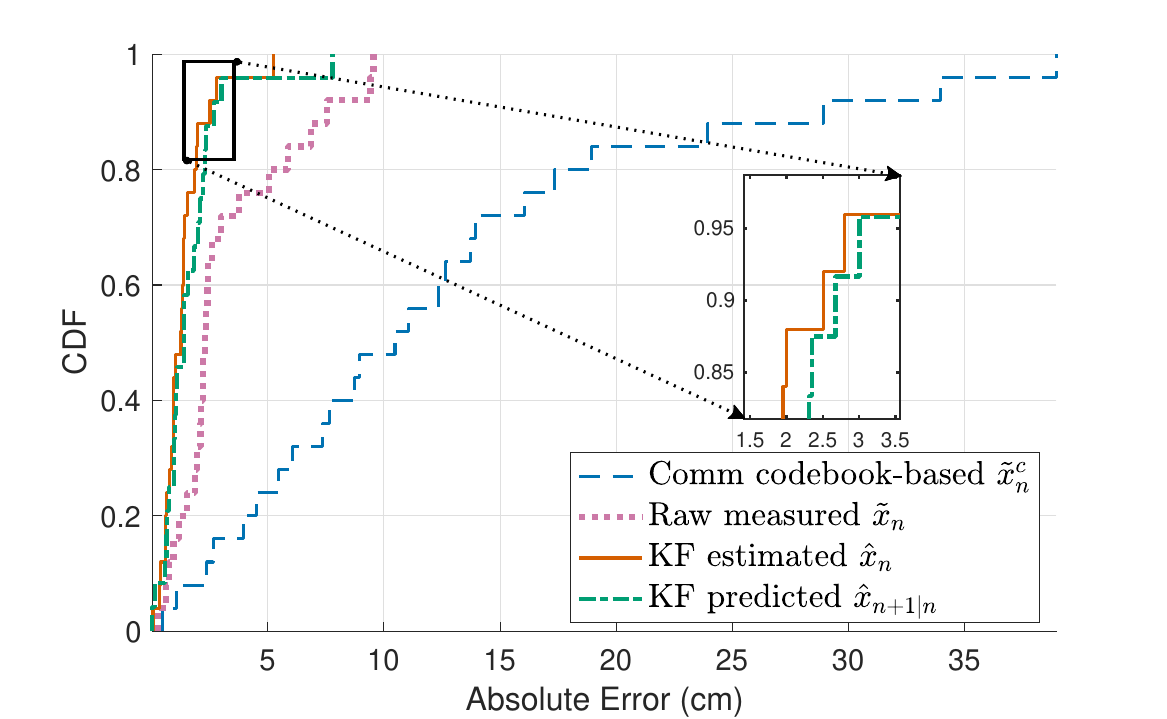}
  \caption{CDF of the absolute cross-range error for codebook-based inference, raw measurement, KF estimation, and KF prediction.}
  \label{fig:range_result}
\end{figure}

Fig.~\ref{fig:range_result} presents the cumulative distribution function (CDF) of the absolute cross-range error. The measured radial distance is first extracted from the echo signal using the leakage component as a timing reference, and then mapped to the cross-range displacement according to the known geometry. For comparison, we also include a communication-only codebook-based estimate. In this approach, the cross-range position is inferred from the center angle of the selected beam, with its estimation accuracy limited by the discrete angular resolution of the codebook.

The raw measured cross-range exhibits fluctuations due to noise and residual multipath effects, while the KF effectively improves estimation and prediction reliability. As shown in Fig.~\ref{fig:range_result}, approximately 90\% of the KF-based cross-range estimates have an absolute error below 3 cm, compared with approximately 30 cm for the communication-only codebook-based inference. This significant improvement provides finer motion information than the discrete beam codebook, enabling more precise beam switching decisions without dedicated sensing waveforms.

\begin{figure}[!ht]
  \centering
  \includegraphics[width=\linewidth]{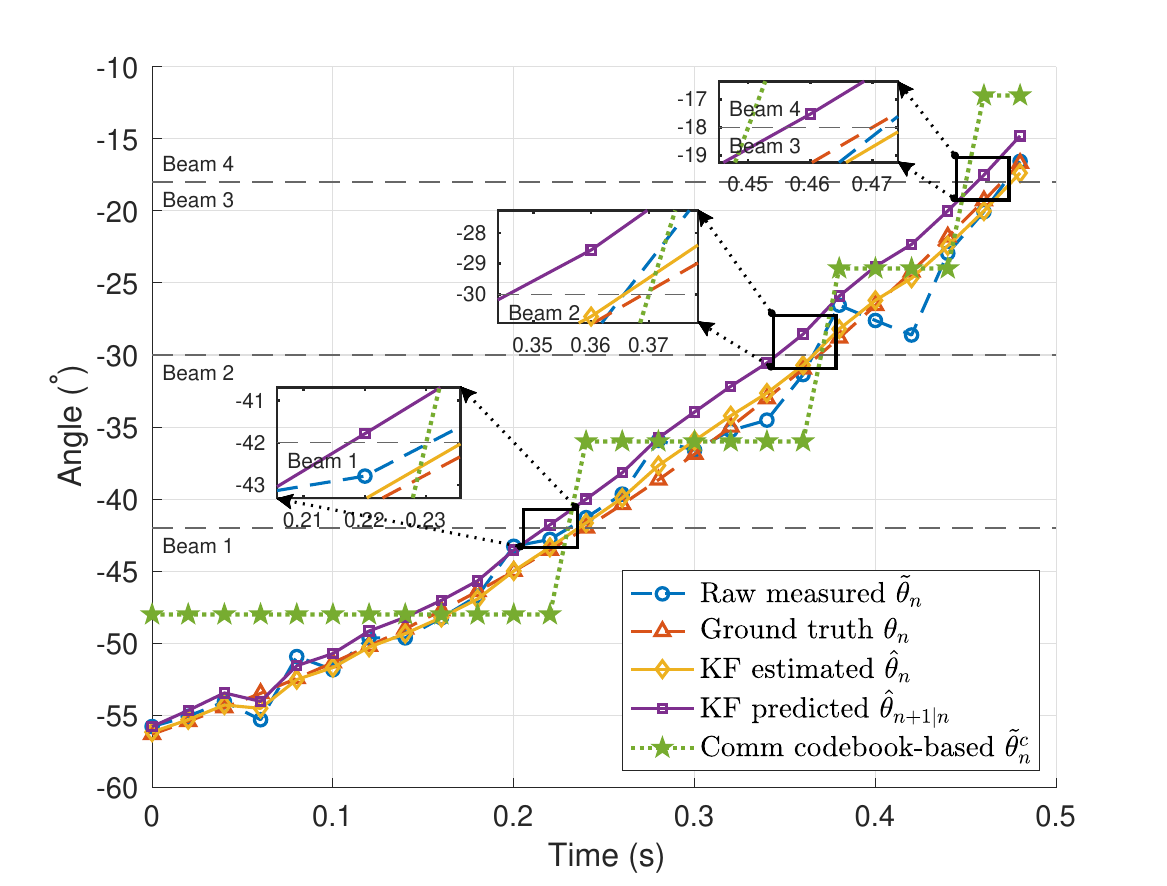}
  \caption{UE angle estimation and prediction with proactive beam switching, compared with communication-only codebook-based angle inference.}
  \label{fig:angle_result}
\end{figure}

Fig.~\ref{fig:angle_result} shows the inferred UE angle obtained from the estimated range, together with the corresponding proactive beam switching behavior and the communication-only codebook-based angle inference. As the UE moves across beam directions, beam switching is triggered when the predicted angle enters the coverage region of a neighboring beam. Compared to the communication-only baseline, which relies on periodic beam sweeping and reactive power comparison, the proposed scheme enables proactive beam switching without dedicated pilot transmission. It is observed that the first two switching instants align with those of the communication-only baseline, while differences appear at the last transition due to the distinct decision mechanisms of the two approaches. Overall, the results demonstrate that accurate motion-aware beam management can be achieved solely based on sensing information extracted from the communication payload signal.

\begin{figure}[ht]
    \centering
    \begin{subfigure}{\linewidth}
        \centering
        \includegraphics[width=\linewidth]{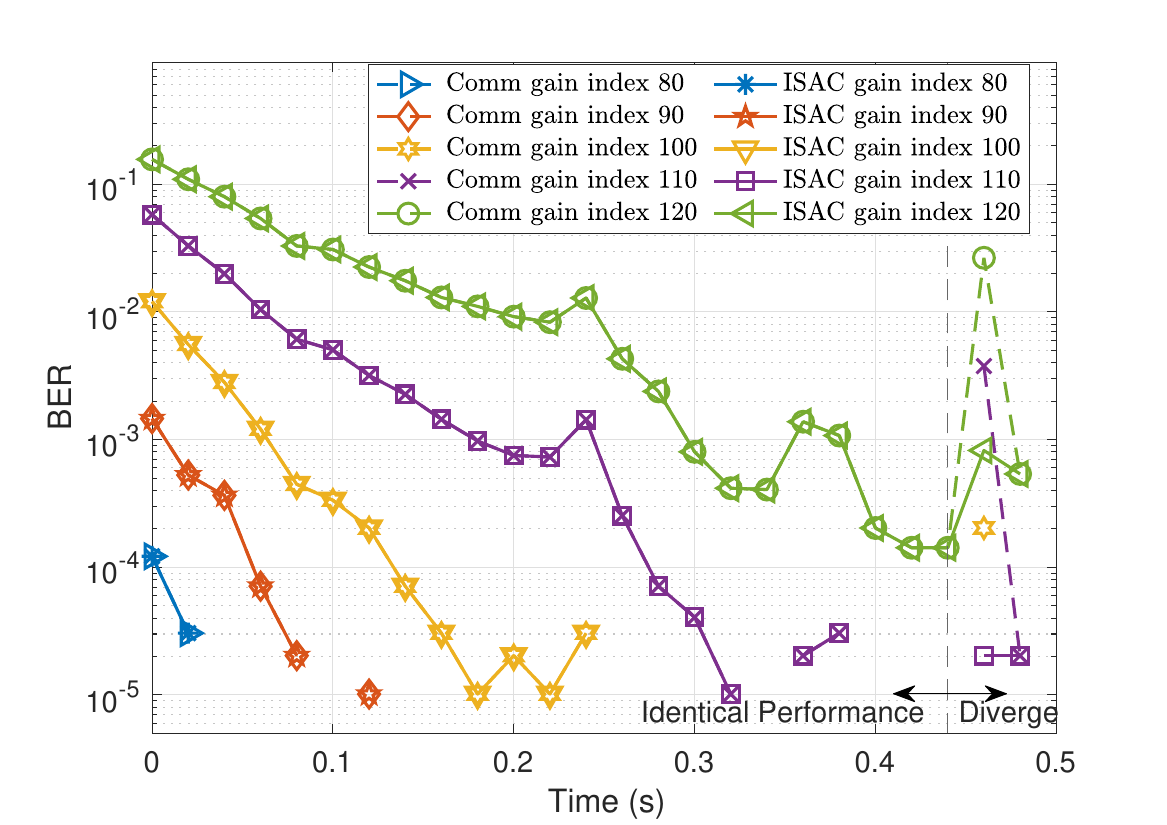}
        \caption{Reliability under beam switching.}
        \label{fig:ber_result}
    \end{subfigure}

    \vspace{0.1cm}
    \begin{subfigure}{\linewidth}
        \centering
        \includegraphics[width=\linewidth]{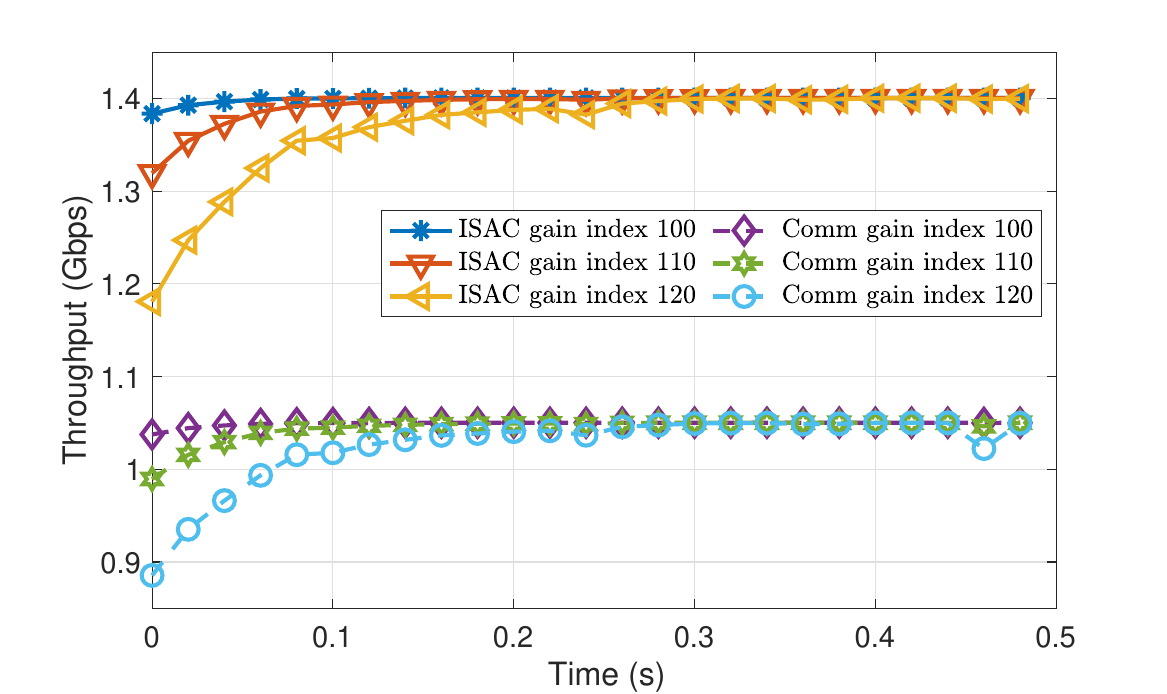}
        \caption{Effective throughput with beam sweeping overhead.}
        \label{fig:tp_result}
    \end{subfigure}
    \caption{Communication performance comparison between the communication-only baseline and the sensing-assisted scheme.}
\end{figure}

Fig.~\ref{fig:ber_result} compares the bit error rate (BER) performance of the communication-only baseline and the proposed sensing-assisted scheme under different transmit gain settings. No forward error correction (FEC) is applied in the current testbed implementation; therefore, the reported BER corresponds to uncoded link performance. In the testbed, a larger gain index corresponds to a lower effective transmit gain. As the UE moves closer to the gNB, reduced path loss leads to a general decrease in BER for both schemes. At higher transmit gains, zero bit errors are observed within certain finite measurement windows, resulting in a measured BER of zero. Since zero values cannot be displayed on the logarithmic scale, the corresponding points are omitted, causing discontinuities in the BER curves. The two schemes exhibit nearly identical performance over most of the trajectory, as the same beams are selected at the first two switching instants. Differences mainly appear at the final switching stage, where distinct beam decisions are made. In this region, the sensing-assisted scheme achieves comparable or slightly improved BER.

Fig.~\ref{fig:tp_result} presents the effective throughput evaluated over 20 ms periods aligned with the NR SS burst periodicity. For the communication-only baseline, 5 ms is assumed to be reserved for beam sweeping, corresponding to 25\% overhead and leaving 15 ms for payload transmission. In contrast, the proposed ISAC scheme avoids this periodic sweeping overhead, allowing the full 20 ms for payload transmission. Consequently, the effective throughput improves from approximately 1.05 Gbps to 1.40 Gbps when BER is sufficiently low, while minor variations arise from BER differences during beam transitions. Together with the accurate cross-range estimation in Fig.~\ref{fig:range_result}, these results show a two-fold benefit of payload-based ISAC: maintaining communication reliability with reduced beam sweeping overhead, while simultaneously providing additional location information.

\vspace{-0.2cm}
\section{Conclusion and Future Work}

This paper presented an experimental validation of payload-based sensing-assisted beam management using 5G NR signals on a 60 GHz mmWave testbed. By reusing communication payload signals for range estimation and integrating geometric inference with Kalman filtering, the proposed scheme enables proactive beam switching without dedicated sensing waveforms or additional pilot resources. Experimental results demonstrated centimeter-level range estimation accuracy and comparable communication reliability to the conventional SSB-sweeping baseline, while improving effective throughput by avoiding periodic beam sweeping under the considered overhead assumptions. These results establish the feasibility of payload-based sensing-assisted beam management on practical mmWave hardware, providing a foundation for further investigation under more complex propagation and mobility conditions.

Future work will extend the proposed framework toward more general mobility patterns and multi-UE scenarios by incorporating angular and Doppler measurements for improved motion prediction and target discrimination. Hybrid or digital beamforming architectures could further provide richer spatial information and support more flexible multi-UE operation. In addition, learning-based methods could exploit historical sensing and communication measurements to predict beam quality variations, anticipate blockage, and facilitate switching to alternative beams. Adaptive sensing and beam management could then adjust measurement and beam selection strategies according to evolving channel conditions.

\vspace{-0.2cm}
\bibliographystyle{IEEEtran}
\bibliography{IEEEabrv,ref}

\end{document}